\documentclass[preprint,12pt, a4paper]{elsarticle}
\usepackage{amssymb}
\usepackage{hyperref}
\usepackage{comment}
\usepackage{svg}
\usepackage{subcaption}
\usepackage{float}
\usepackage{enumitem}
\usepackage{soul}

\hypersetup{
    colorlinks=true,
    linkcolor=blue,
    filecolor=magenta,      
    urlcolor=cyan,
}

\journal{SoftwareX}

\begin{document}
\renewcommand{\labelenumii}{\arabic{enumi}.\arabic{enumii}}

\begin{frontmatter}
 


\title{CitySolution: A complaining task distributive mobile application for smart city corporation using deep learning}

\author{Farhatun Shama}
\author{Abdul Aziz\corref{cor1}}
\ead{abdulaziz@cse.kuet.ac.bd}
\author{Lamisa Bintee Mizan Deya}

\cortext[cor1]{Corresponding author}
\address{Department of Computer Science and Engineering, Khulna University of Engineering \& Technology,  Khulna 9203, Bangladesh}

\begin{abstract}

The lack of an automated online platform for reporting citizens' complaints, coupled with the city corporations' struggles in managing them, presents significant challenges. Furthermore, the availability of resources is very limited to higher authorities for monitoring progress. The primary objective of this paper is to develop two Android applications and to categorize complaints automatically using a deep learning model created on the Teachable Machine. With the citizen-oriented application, individuals can easily report complaints by capturing pictures of their municipal issues. The authority version of the application provides categorized complaints, along with location and status details. Higher authorities can monitor the municipal progress, thereby enhancing transparency, and efficiency and promoting smart city development on a nationwide scale.

\end{abstract}


\end{frontmatter}


\section{Motivation and significance}

This paper is driven by the need to rectify the inadequacies in the current complaint management and categorization system. Citizens frequently encounter challenges in reporting issues and monitoring the progress of complaints due to the absence of a seamless online platform. At the same time, city corporations struggle with the classification and management of reported problems, resulting in delays. Additionally, higher authorities lack efficient mechanisms for monitoring city corporations' management of reported problems. Consequently, numerous problems remain unaddressed, while others are resolved belatedly. These highlight the indispensable need for enhanced systems and technologies in civic governance.
\section*{Metadata}
\label{}

\begin{table}[H]
\begin{tabular}{|l|p{6.5cm}|p{6.5cm}|}
\hline
\textbf{Nr.} & \textbf{Code metadata description} & \textbf{Please fill in this column} \\
\hline
C1 & Current code version & v1 \\
\hline
C2 & Permanent link to code/repository used for this code version &\url{https://github.com/Shama-33/CitySolution.git}  \\
\hline
C3  & Permanent link to Reproducible Capsule & None\\
\hline
C4 & Legal Code License   & Apache-2.0 license \\
\hline
C5 & Code versioning system used & git \\
\hline
C6 & Software code languages, tools, and services used & Android Studio, Java, Firebase, Teachable Machine by Google \\
\hline
C7 & Compilation requirements, operating environments \& dependencies & Android Studio Dolphin Version 2021.3.1\\
\hline
C8 & \ If available Link to developer documentation/manual& \url{https://github.com/Shama-33/CitySolution/blob/main/User%20Manual.pdf}
\\
\hline
C9 & Support email for questions & farhatunshama@gmail.com
\newline lamisa.deya2001@gmail.com 
\newline abdulaziz@cse.kuet.ac.bd
\\
\hline
\end{tabular}
\caption{Code metadata (mandatory).}
\label{codeMetadata} 
\end{table}

In response to these problems, the paper aims to provide a user-friendly Android application that allows citizens to report a wide range of issues. The precise classification of the issues by the embedded deep learning model will give city authorities a methodical way to handle each issue. Within the application, several graphical representations will provide an overview of the distribution of problems across different categories, insights into the status of reported issues, and a breakdown of reported problems based on the cities they originate from. Thus, higher authority can effectively track and evaluate the status of the city corporation's tasks. The applications also place a high priority on user privacy, guaranteeing that private data is kept private and protected. The system leverages Firebase's robust security protocols to effectively manage and protect user data. The applications utilize comprehensive security features offered by Firebase, including secure data storage, SSL-encrypted data transmission, and periodic security assessments, to ensure that user data is handled securely and with great care. Furthermore, Firebase’s email authentication is used to verify users’ identities before registration, adding an additional layer of security. To enhance the integrity of the authority version of the app, a QR code scanning feature has been introduced. This feature is employed for city corporation employee registration, ensuring that only authorized personnel can register and operate the application. Central admin can generate these QR codes through the app using the information of ID, first name, last name, and city. Authorities can use these QR codes to register. While registering, the app checks if the QR code’s information is stored in Firebase. If anyone creates a similar pattern of the QR code provided by the higher authority and tries to register, they cannot proceed because the QR code is not provided by the higher authority. The higher authority maintains a node in Firebase to store the authority's info, preventing unauthorized access and maintaining the authenticity of actions taken within the app, such as updating or resolving complaints. Moreover, to provide an inclusive and accessible experience for all users, a dual-language feature has been implemented. This enhances usability by accommodating diverse linguistic preferences, ensuring that individuals from varied backgrounds can interact with the platform seamlessly. The user can also see feedback from the city corporation.

The incorporation of fused location services by Google offers a dual functionality that enables automatic location retrieval for users while providing them with the option to manually select their desired location. This feature enhances user convenience by eliminating the need for manual input when reporting complaints. Furthermore, users can report issues even in the absence of internet connectivity or later by manually inputting the location details. Authorities receive precise location information that can be effortlessly redirected to Google Maps for visualization. Additionally, authorities can communicate directly with users via email to obtain further information about the reported problem. The authorities have the power to correct the category of a complaint if the deep learning model misclassified it. Furthermore, the authorities can switch fraudulent reports to a unique category labeled as 'fake complaints'. This extensive range of features not only simplifies the reporting process but also improves the accuracy and effectiveness of complaint management in city corporations.

The major contributions of this research paper include the following:
\begin{enumerate}[label=\textit{\roman*}]

\item Two mobile applications have been developed for the purpose of automating the complaint task distribution by incorporating a deep learning-based model.
\item A benchmark image dataset comprising 3,000 images has been prepared for the training of the model that contains four classes- “Damaged road”,” Flood”,” Trash” and “Homeless people”.
\item A deep learning model is integrated using Google’s Teachable Machine, which automates categorization and reduces time and the chance of human error.
\item Fused location services are integrated with the application that automatically fetches the location of the user.
\item Redirection to email and map gives authorities precise location information and enables direct communication with users for further details.
\item Several graphical representations have been integrated to provide an overview of the progress of complaints, including city-wise, acknowledged, pending, and solved issues, as well as category-wise distribution.
\item Dual language support has been included to improve accessibility and user experience.
\item Central admin can generate QR codes through the app using the info of ID, first name, last name, and city. Authorities can use these QR codes to register.

\end{enumerate}

Several applications have been developed for the same purpose, such as the complaint management system \cite{one}. In this existing web application, citizens must manually select the category and location of the problem to file a complaint or notify the authorities about issues in their area. They must then provide a written description of the problem. The authority can view the list of these problems and provide feedback. However, this system is not automated and is designed for use by a specific organization or municipality. All problems, even those from locations outside the city boundary, are displayed to the authority, making it difficult to filter out irrelevant data. Additionally, since the complaints are text-based, it is challenging to verify their authenticity. Moreover, there is no centralized monitoring system to ensure that city corporation employees are fulfilling their duties properly.
Municipal Complaints Unit is another mobile application that allows citizens to submit complaints, which can be viewed and addressed by the relevant authorities \cite{two}. In this system, users manually select the category and location of the complaint, along with a description and an image. The authority is then able to view this information and update the complaint status upon resolution. However, this system is only applicable to a single city, limiting its use to the relevant authority of that city. Additionally, there may be issues with irrelevant data, as users may submit complaints about locations outside of the city boundary. The manual selection of categories by the user may also lead to errors and inconvenience.

A crowdsourcing software system \cite{1}, has been designed to help small municipalities efficiently collect data on public infrastructure conditions. The system relies on residents reporting various issues, which are then communicated to local governments for better decision-making and proactive management. Unlike the proposed approach, which employs a Deep Learning model to automatically classify and analyze the reported issues, their system focuses solely on gathering and organizing user-generated reports.


 Google’s Teachable Machine leverages transfer learning, enabling it to attain an accuracy rate of up to 100\% as reported in \cite{3}.  The research in \cite{4} utilized transfer learning to fine-tune the parameters of the pre-trained VGG19 network for image classification. 

In \cite{8},  an application integrates key city services like event listings, public transport information, digital library access, health care resources, and feedback systems. By providing these services through a mobile platform, the application strengthens communication between citizens and city authorities, promoting greater engagement and efficient management of urban services. This work demonstrates how mobile technology can make city services more accessible and encourage active citizen participation in city governance. How mobile apps can improve smart cities, citizen satisfaction, and involvement is discussed in the paper \cite{9}. It focuses on the elements that affect how these apps are used and accepted, offering insightful information on user preferences and technology acceptability. The study provides important recommendations for enhancing municipal services through technology by carefully analyzing the effects of user interface design and service integration on user adoption rates. The paper \cite{10} explores how IoT technologies might be integrated into smart grid applications, emphasizing developments in smart energy metering, home automation, and bidirectional power flow. The potential of IoT technology to revolutionize urban energy management is demonstrated by this study. It talks about the difficulties in putting them into practice and offers alternatives that might result in more sustainable and effective patterns of energy use. Research \cite{11} discusses how grounded theory is applied in the creation of mobile applications intended for smart cities. The results provide a solid foundation for developers to create applications that are efficient and easy to use, improving the livability of cities. The impact of these apps on public service accessibility as well as the significance of community input in the design process are also highlighted in the study. In \cite{12}, the design of a kit for smart cities with the goal of simplifying the development of mobile and online applications is investigated. This kit offers necessary resources and instructions that speed up the implementation of smart city solutions and encourage developer creativity. The study goes into detail on the kit's modular design, which enables scaling and customization to meet the demands of various urban environments.

This system employs a deep learning model that was developed with the help of the Teachable Machine by Google that categorizes the complaints into "Damaged Road," "Flood," "Trash," or "Homeless People". Teachable Machine only requires a browser and a dataset to generate a model. It uses deep learning, convolutional neural networks (CNN), MobileNets, ImageNet, Transfer Learning, etc. The term deep learning or deep neural network refers to artificial neural networks with multiple layers that can handle large amounts of data and outperform classical methods in several fields \cite{three}. CNN is one of the most popular deep learning networks that contains a convolutional layer. It has excellent performance in deep learning problems, especially in applications that deal with image data \cite{three}. Transfer learning is a deep learning technique where a model is trained and developed for one task and then it is tweaked and reused for a second related task. This base model is called the pre-trained model. The dataset for the second related task is usually smaller than the one used to train the pre-trained model \cite{four}. MobileNets is a class of lightweight and efficient CNNs that are used for mobile and embedded vision applications \cite{ashikuzzaman2021danger, five, tithy2021deep}. Due to the lightweight and compact design of MobileNets, they are well-suited for being the pre-trained model for transfer learning \cite{fime2024audio}. This model is mainly for image classification and detection. ImageNet is a hierarchical image database that contains millions of cleanly labeled, full-resolution, sorted images. This organizes the images into 1000 distinct categories \cite{six}. In Teachable Machine, “Supervised Learning” and “Transfer Learning” are implemented. As the pre-trained model, the 'Teachable Machine' uses MobileNets, which is trained with the images of ImageNet. So, the models trained by users show high accuracy even with relatively smaller-sized datasets. MobileNets is built from depth-wise separable convolutions. The reduction in parameter passing allows it to be more compact and provides it with the ability to run on mobile devices. Transfer learning allows us to fine-tune the pre-trained model (MobileNets) to recognize the new data. So, there was no need to start from scratch with random weight initiation. Since the pre-trained model already recognized a lot of data, the deep learning model had a high accuracy, precision, recall, and F1 score. 


\section{Software description}

\begin{figure}[!b]
    \centering
    \includegraphics[width=0.8\textwidth]{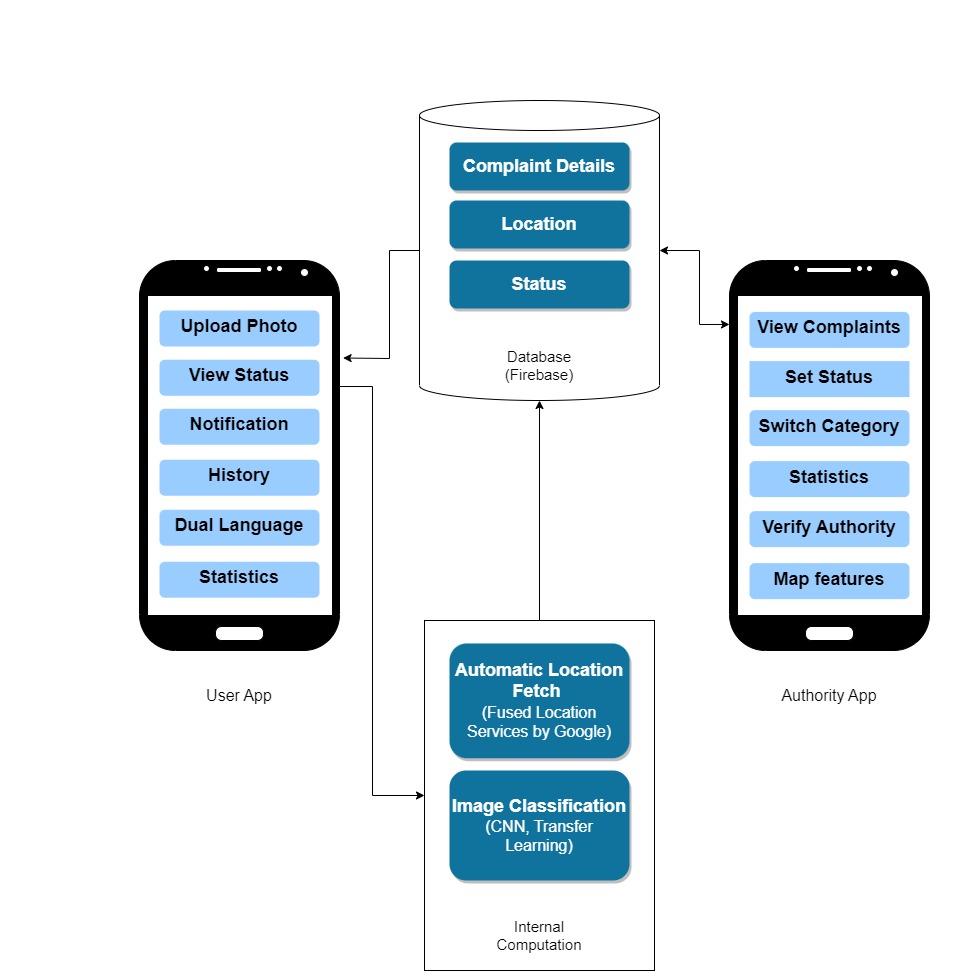}
    \caption{The system overview.}
    \label{fig:1}
\end{figure}

\subsection{Software architecture}
This software consists of two Android applications. One for the user (the citizens) and the other for the authorities (both central and city corporation authorities). The following tools were used for the creation of this system.
\begin{enumerate}[label={\alph*.}]
    \item Android Studio: An integrated development environment (IDE) designed for developing Android applications, offering a user-friendly interface and a comprehensive toolkit that is compatible with Windows, macOS, and Linux operating systems. It provides features such as code completion, debugging tools, and built-in support for version control.
    Android Studio and native Android development have been chosen to ensure optimal performance and access to device-specific features, such as fused location services and Google Maps integration. Native development offers a more responsive and reliable user experience compared to multi-platform frameworks like React Native.
    \item Java: A widely used, object-oriented programming language that is known for its reliability, portability, and large developer community support, making it an ideal choice for Android app development. It offers platform independence, strong memory management, and a vast ecosystem of libraries and frameworks.
    \item Firebase: A cloud-based database and computing service provided by Google, that offers features like authentication, real-time databases, and Firebase storage. It is chosen for its security, ease of use, and free access. Firebase also provides analytics, crash reporting, and cloud messaging capabilities.
    Firebase is selected over other database solutions due to its robust real-time capabilities, seamless integration with Android applications, and built-in authentication features, which simplify user management and data synchronization.

    \item Fused Location Services: A part of Google Play Services, which offers location information from various sensors. It is used in the application to fetch location data, which is viewable via the Google Maps app. Fused Location Services provides accurate and battery-efficient location updates, making it suitable for location-aware applications.
    \item Teachable Machine: A web-based tool developed by Google for creating deep learning classification models. It uses the Tensorflow.js library for training the models in the browser, allowing for deep learning model training even with limited hardware resources \cite{seven}. It also generates compact models that are executable on mobile devices \cite{eight}. Teachable Machine simplifies the process of training machine learning models by providing a user-friendly interface and pre-trained models for quick experimentation.
    The Google Teachable Machine has been selected due to its ease of use, quick development and deployment capabilities, and efficient performance on mobile devices. It leverages MobileNets, which are lightweight and suitable for mobile applications, providing high accuracy with minimal computational resources. This choice allows to implement a powerful yet efficient deep-learning model within the application.

\end{enumerate}
The architecture of the system can be broadly classified into three stages. The development of the deep learning model, the development of the user version of the “CitySolution” application, and the development of the authority version of the “CitySolution” application. The overall system overview consists of the steps in Fig. \ref{fig:1}.

\subsubsection{Developing the deep learning Model}
The steps involved in developing the deep learning model are:
\begin{enumerate}[label={\alph*.}]
    \item {Dataset Collection:} A benchmark has been prepared for developing the applications. The four classes contain 5494 images in total. Individually, the "Damaged Road", "Flood", "Trash" and "Homeless People" classes contain 1072, 1183, 1616, and 1623 images respectively. The images were collected by manually capturing the pictures using the camera of the mobile phone as well as from various sources including the Internet. Then, these four categories of data were uploaded to the Teachable Machine. The Teachable Machine performed some preprocessing of the data. The preprocessing involved - resizing each image to a square to ensure that the inputs are of the same size and dimensions. It also typically normalizes the images to fit into the [0,1] range. Furthermore, augmentation techniques are applied to increase variability.  
    
    \item {Transfer Learning:} The Teachable Machine employs transfer learning and supervised learning techniques to train the model within the user's browser. This is achieved using MobileNets, a pre-trained model that has been trained on ImageNet images. Based on this pre-trained model, a custom model was trained. An epoch of 200, a batch size of 16, and a learning rate of 0.001 were used to train the model. From each category of the problems, 85\% images were used as training data, and the rest were used as testing data. The testing data for Damaged Road, Flood, Trash, and Homeless People were 161, 178, 243, and 244 respectively. The correctly classified data were 160, 174, 239, and 240 for the previously stated order of classes.
    
    \item {Exportation of the Model:} Following training, the model was directly exported from the Teachable Machine in tflite format and integrated into the user version of "City Solution" application. This tflite format facilitates seamless deployment of the model in mobile applications.
\end{enumerate}

\subsubsection{Developing the “CitySolution” Application} 
The CitySolution's user application and the authority application employ basic features such as login, sign up, user profile, notifications, etc. It also facilitates language, security, and profile settings. The user can view the list of complaints he or she submitted as well as the current status of the problem using the user application. The authority application has two major segments such as one for city corporation employees and the other for higher authority. The city corporation employees can view all the classified complaints within their city and can update the status of the problem. On the other hand, the higher authority can monitor any city i.e can see all the problems as well as remove employee accounts. The CitySolution's user application has a deep learning model embedded in it.  The working flow diagram of both the user application and the authority application are represented in Fig. \ref{fig:2}, and Fig. \ref{fig:3}, respectively.  

\begin{figure}[H]
    \centering
    \includegraphics[width=0.8\textwidth, height=0.68\textwidth]{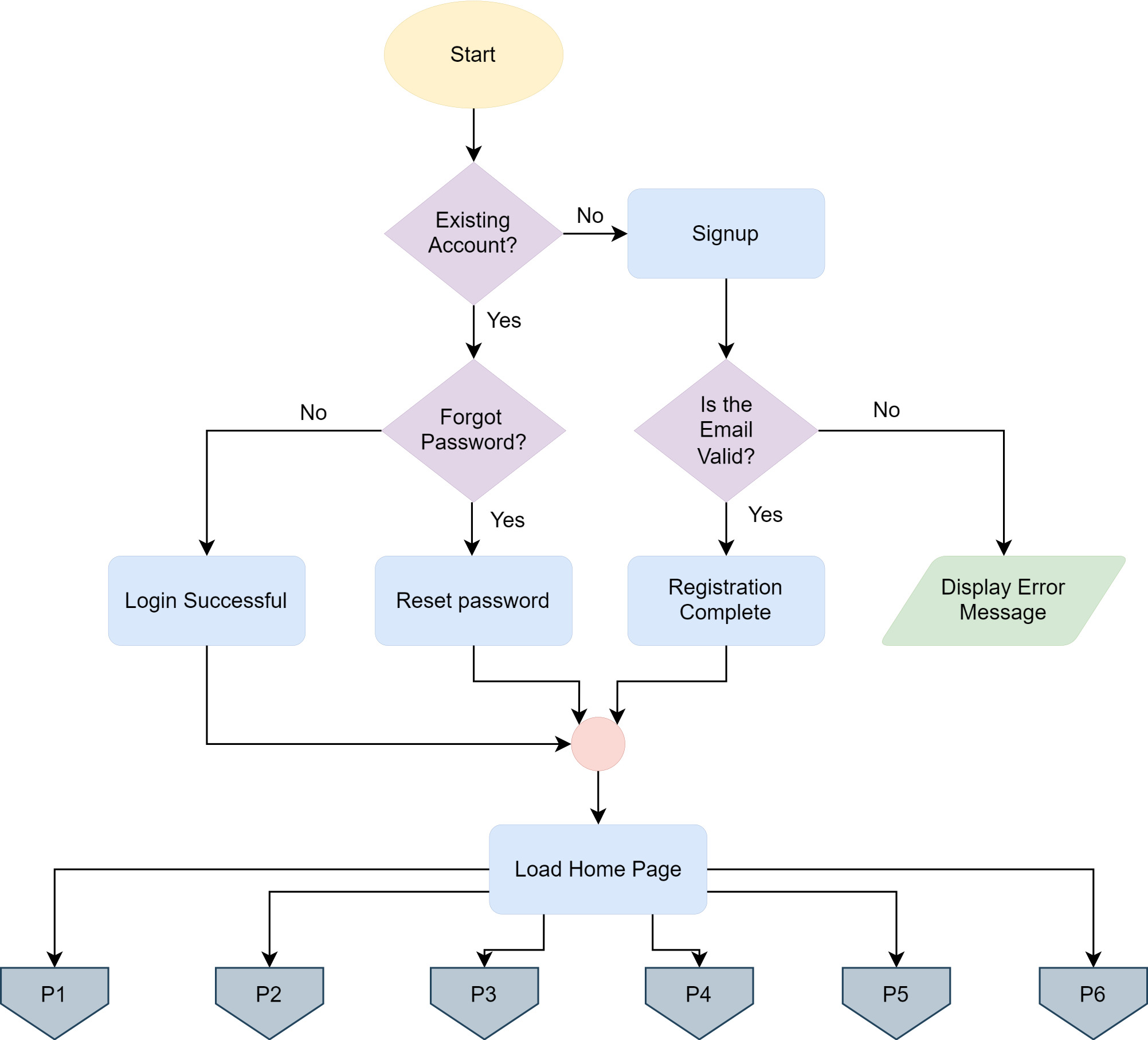}
    \caption{The working flow diagram of CitySolution's user version.}
    \label{fig:2}
\end{figure}

\begin{figure}[H]
    \ContinuedFloat
    \centering
    \includegraphics[width=0.8\textwidth]{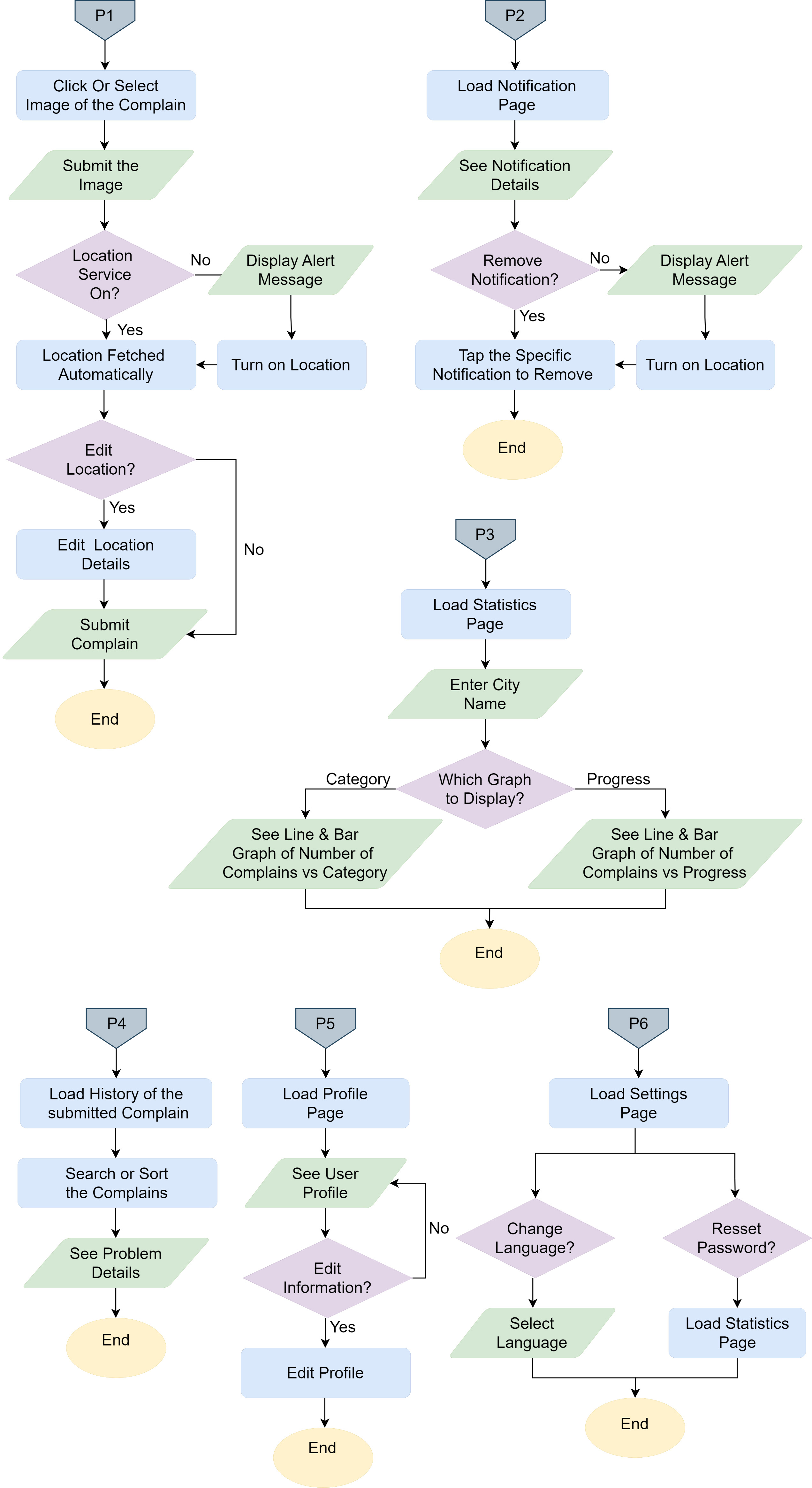}
    \caption{The working flow diagram of CitySolution's user version (cont'd).}
    \label{fig:2}
\end{figure}

\begin{figure}[H]
    \centering
    \includegraphics[width=\textwidth]{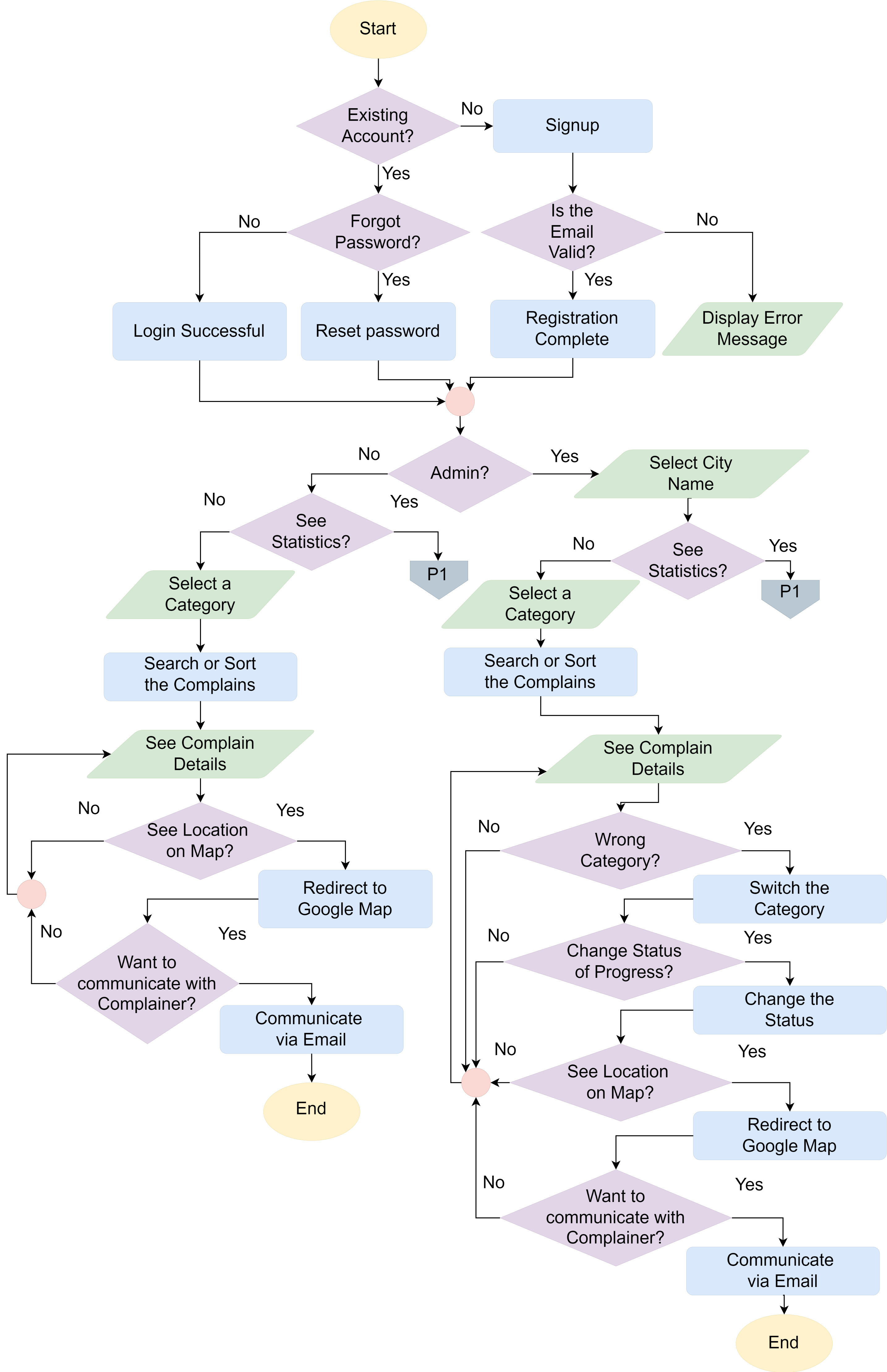}
    \caption{The working flow diagram of CitySolution's authority version.}
    \label{fig:3}
\end{figure}
\begin{figure}[H]
    \ContinuedFloat
    \centering
    \includegraphics[width=\textwidth]{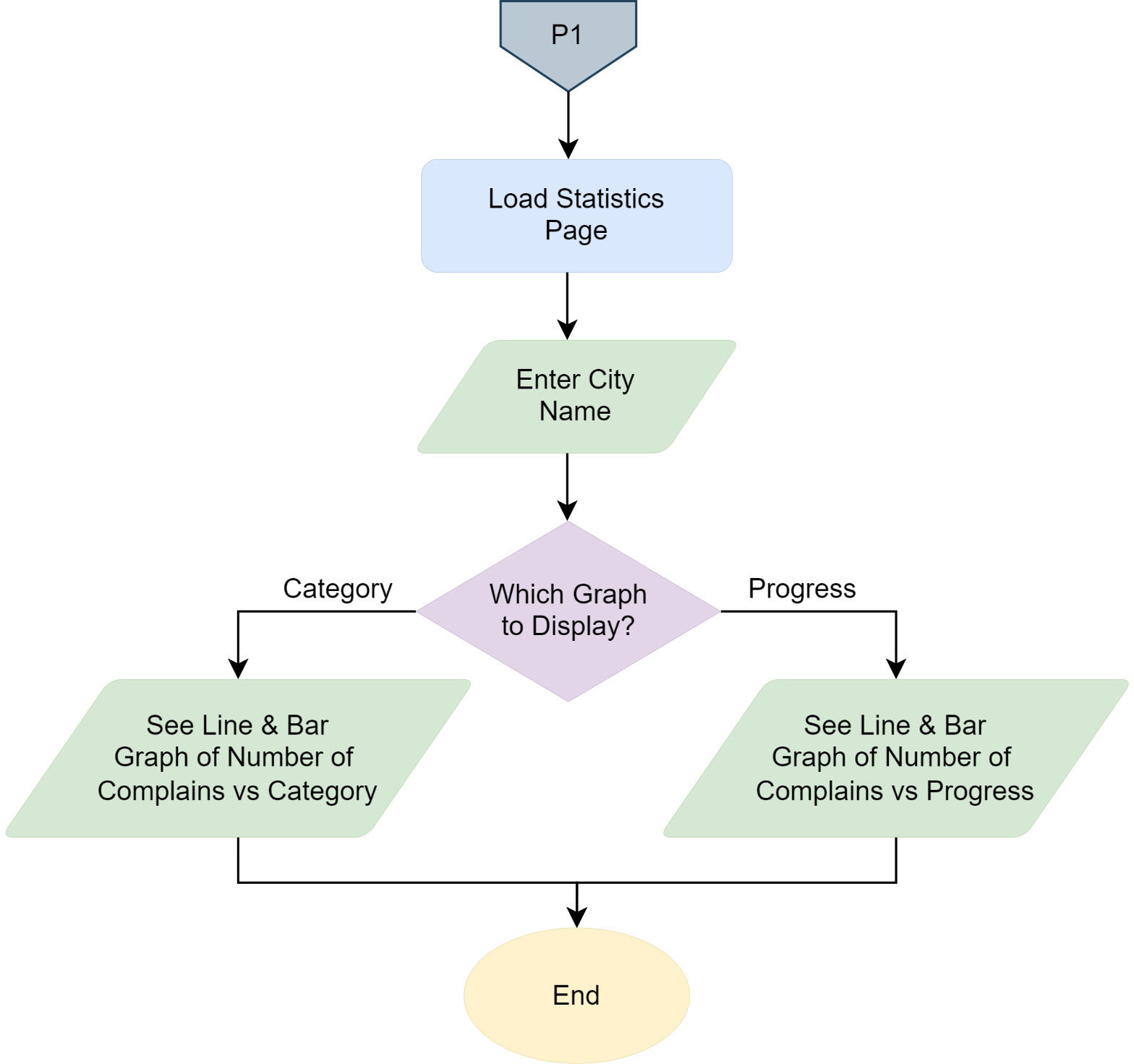}
    \caption{The working flow diagram of CitySolution's authority version (cont'd).}
    \label{fig:3}
\end{figure}

 \subsection{Software functionalities}
The software provides many user-friendly functionalities. The location of the problem is fetched automatically by the fused location services. The complaints are classified automatically with the help of a deep learning model which saves valuable time and resources. The user receives in-app notifications on status updates and also gets feedback from the city corporation. The city corporation can view the problem location with the help of Google Maps. Graphical representation has been added for transparency and easy monitoring. The citizens and higher authority can view the graph of any city within the country and view the ratio of the complaints that are unaddressed, solved, or under process. They can also view the number of complaints in each category. This promotes transparency and provides valuable insights for urban planning. There is also a dual language feature for easy use.

\section{Illustrative examples}

\begin{figure}[H]
    \centering
    \begin{subfigure}[b]{0.3\textwidth}
        \centering
        \includegraphics[height=6cm]{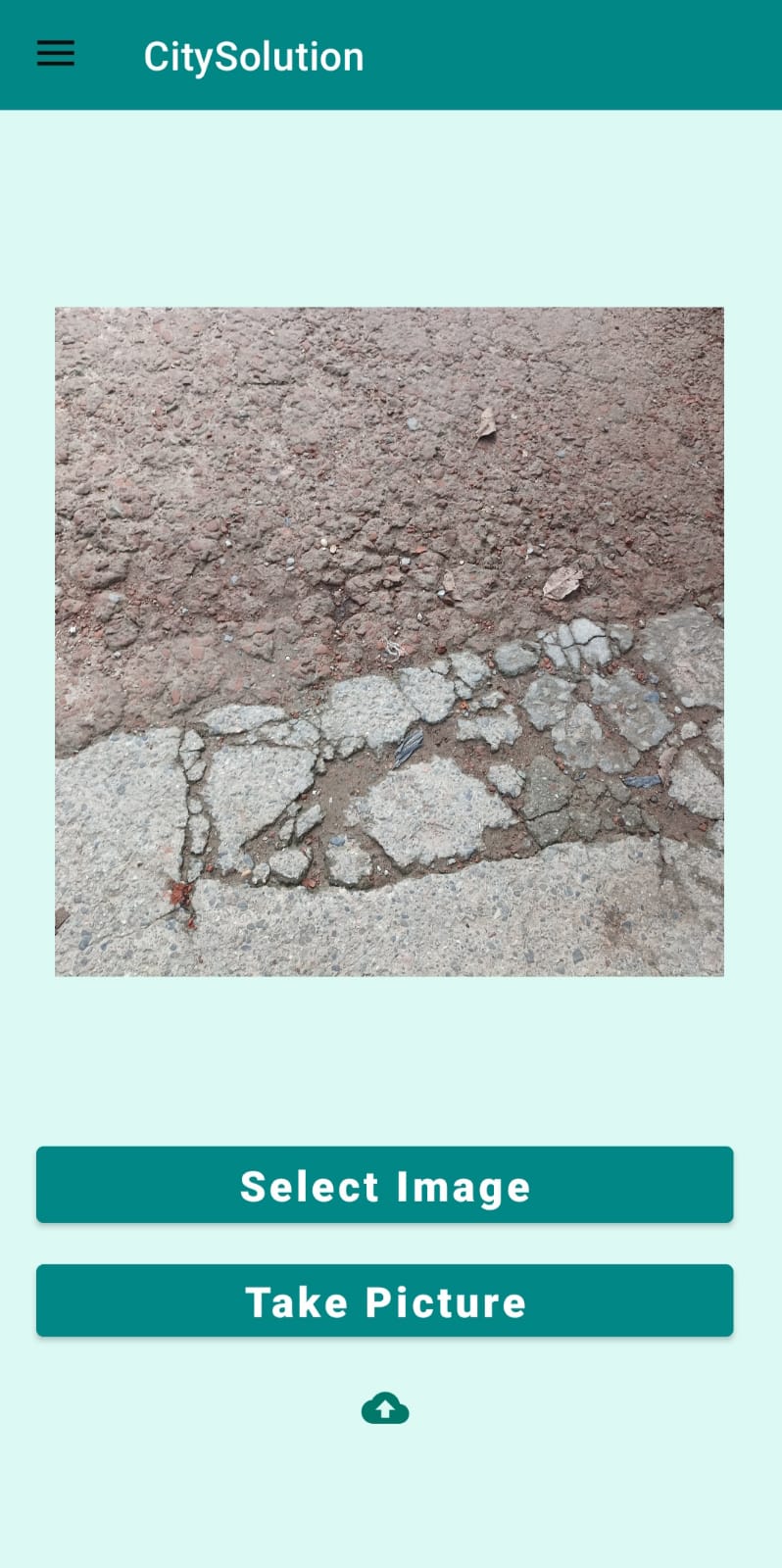}
        \caption{Upload Complain}
        \label{fig:sub1}
    \end{subfigure}
    \begin{subfigure}[b]{0.3\textwidth}
        \centering
        \includegraphics[height=6cm]{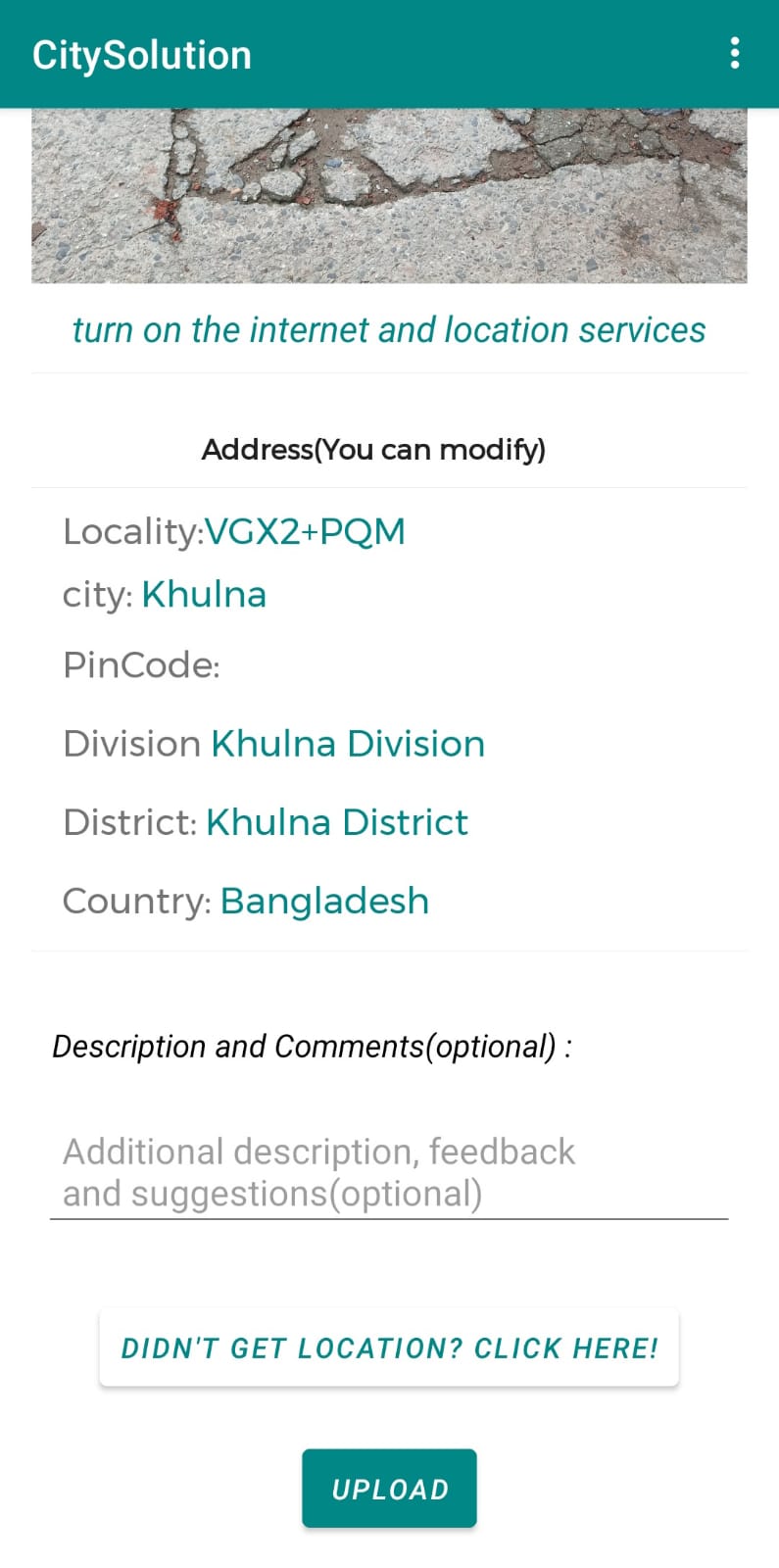}
        \caption{Automatic Location Fetch}
        \label{fig:sub2}
    \end{subfigure}
    \begin{subfigure}[b]{0.3\textwidth}
        \centering
        \includegraphics[height=6cm]{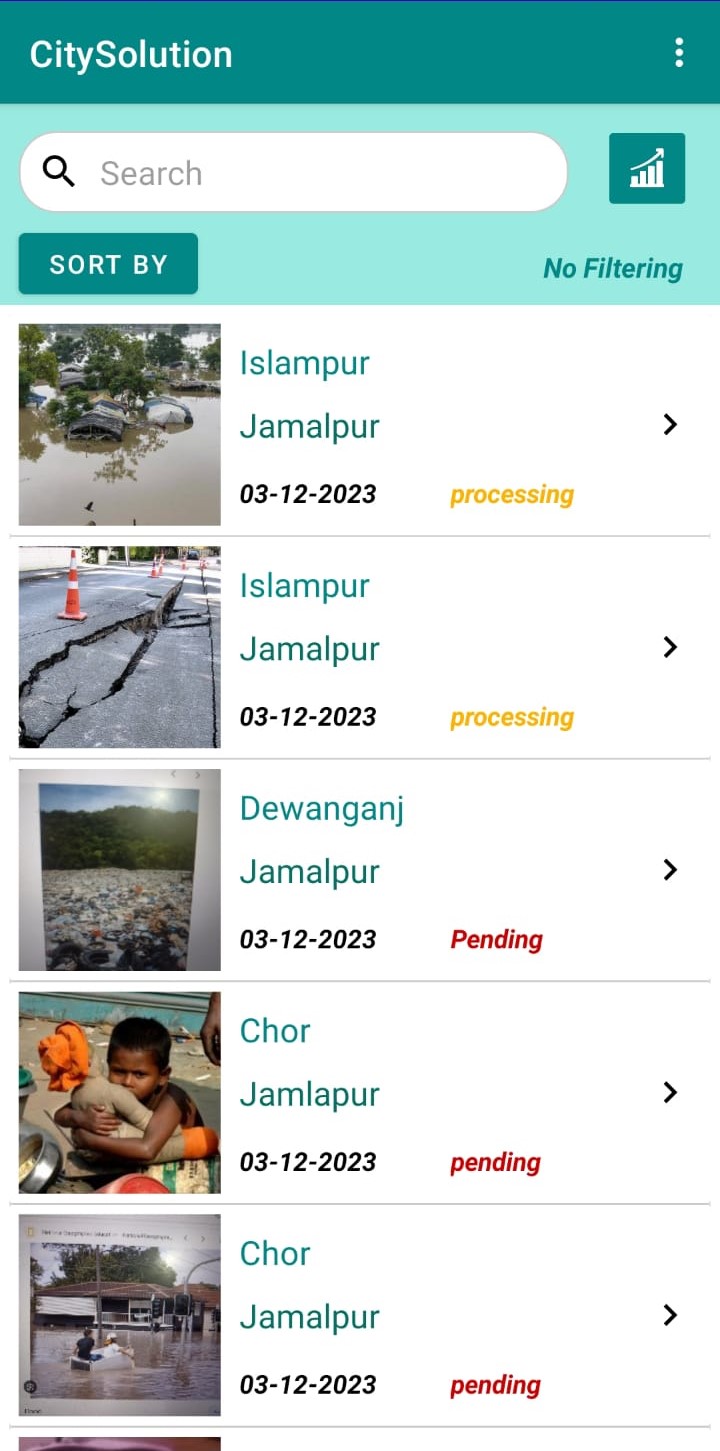}
        \caption{History}
        \label{fig:sub3}
    \end{subfigure}
    
    \medskip 
    
    \begin{subfigure}[b]{0.3\textwidth}
        \centering
        \includegraphics[height=6cm]{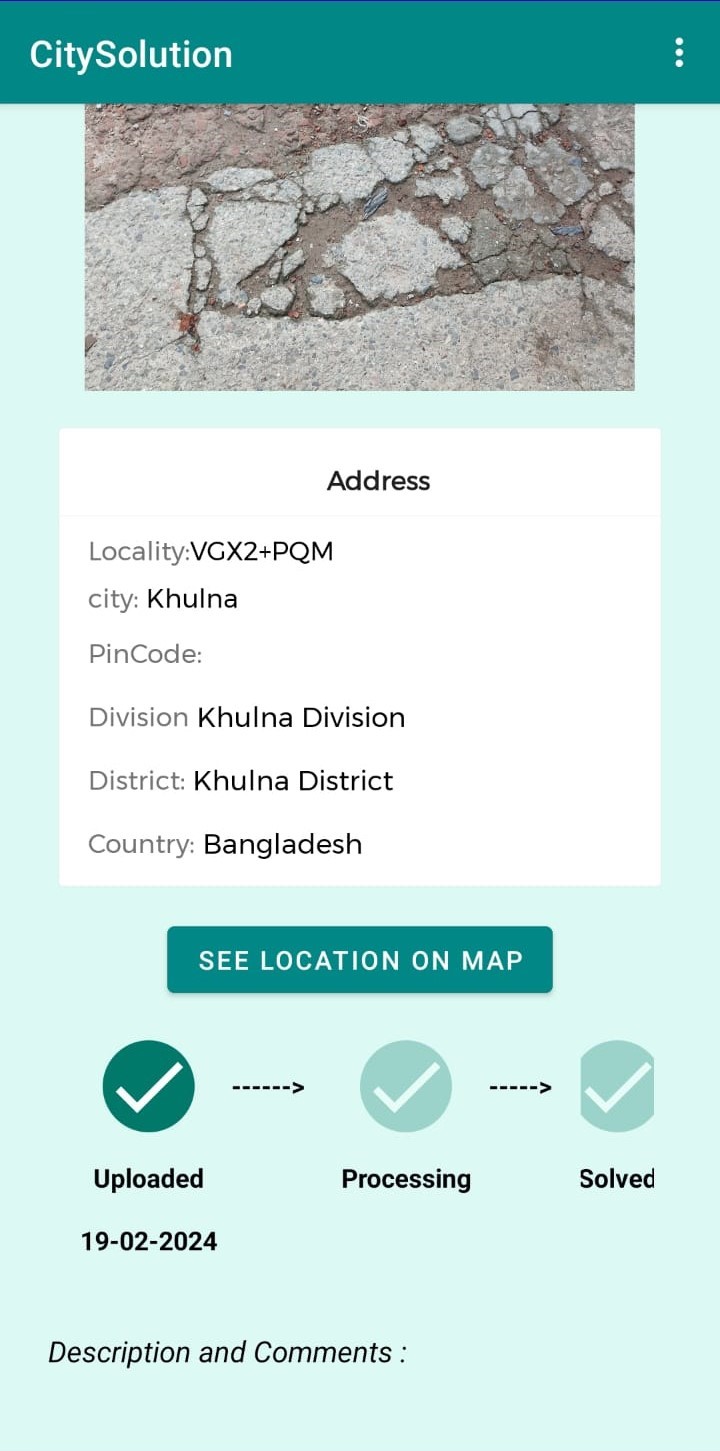}
        \caption{History Detail}
        \label{fig:sub4}
    \end{subfigure}
    \begin{subfigure}[b]{0.3\textwidth}
        \centering
        \includegraphics[height=6cm]{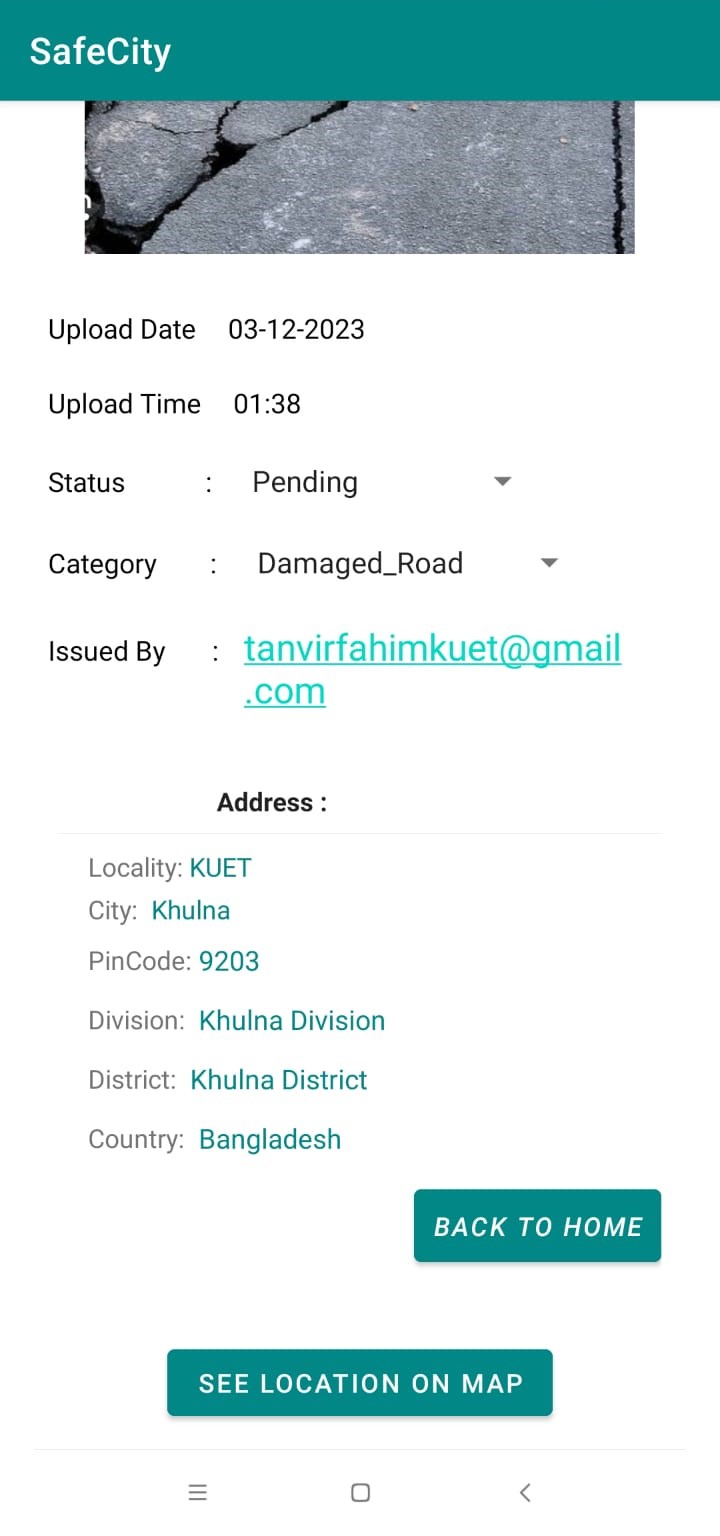}
        \caption{Problem Detail}
        \label{fig:sub5}
    \end{subfigure}
    \begin{subfigure}[b]{0.3\textwidth}
        \centering
        \includegraphics[height=6cm]{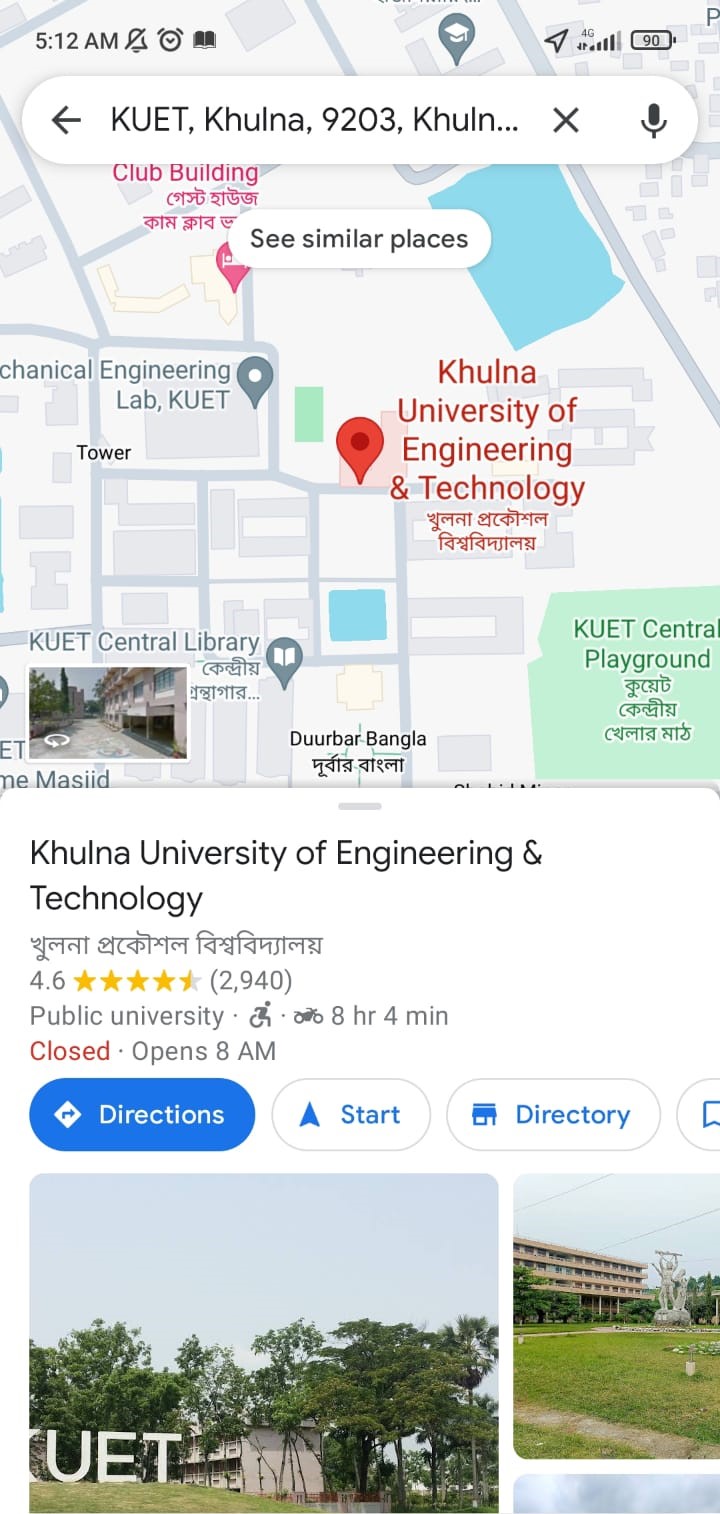}
        \caption{Redirect to Google Map}
        \label{fig:sub6}
    \end{subfigure}
    
    \medskip 
    
    \begin{subfigure}[b]{0.3\textwidth}
        \centering
        \includegraphics[height=6cm]{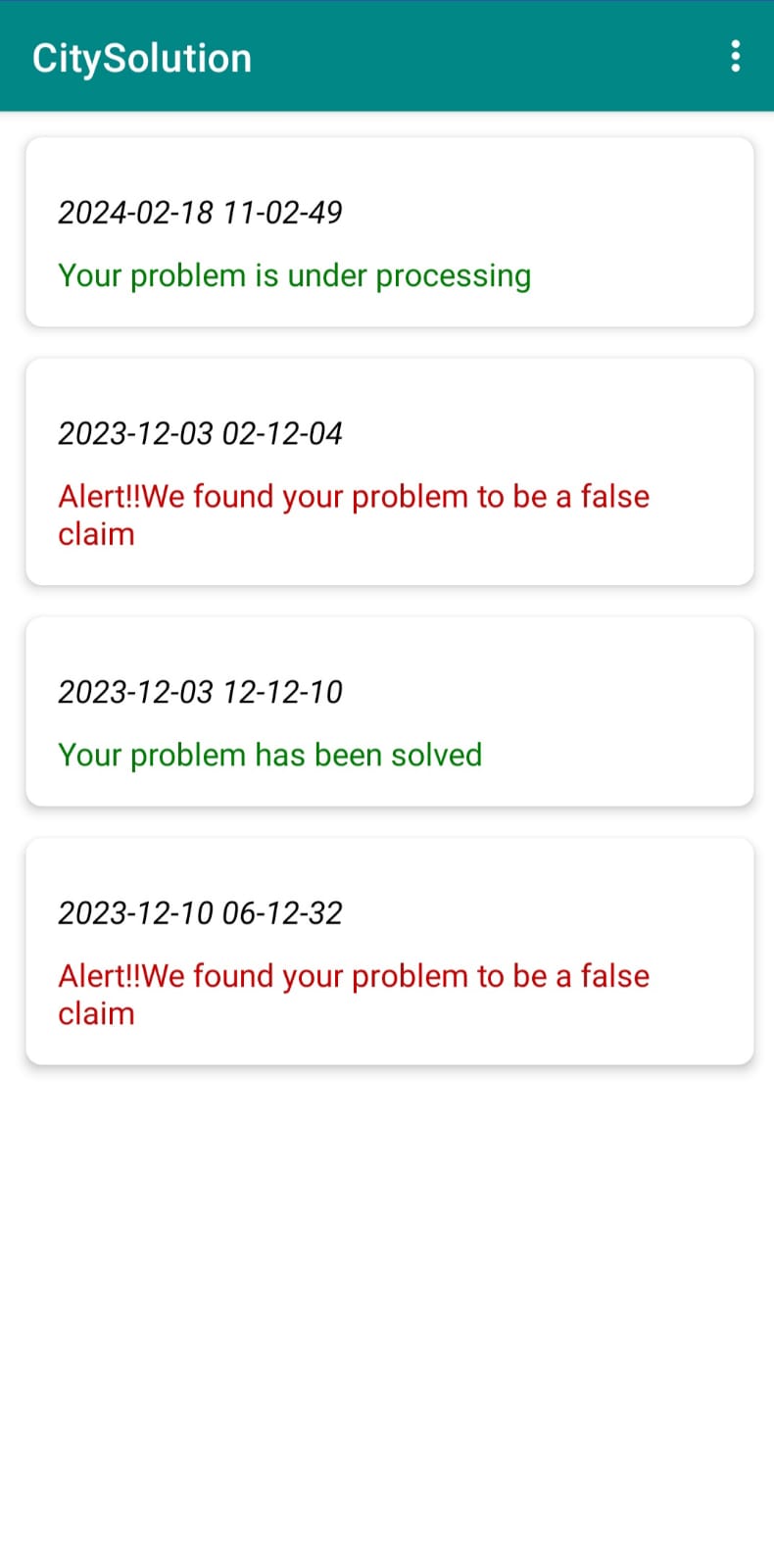}
        \caption{Notification}
        \label{fig:sub7}
    \end{subfigure}
    \begin{subfigure}[b]{0.3\textwidth}
        \centering
        \includegraphics[height=6cm]{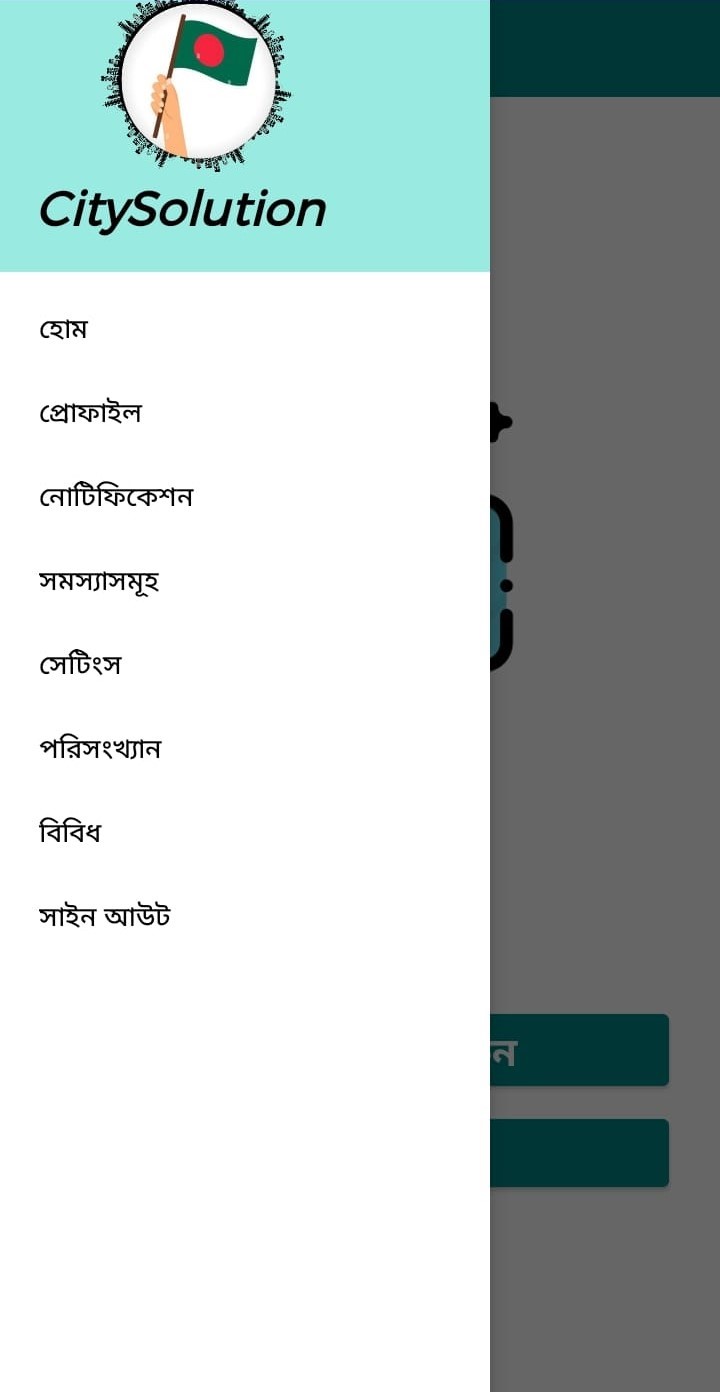}
        \caption{Dual Language}
        \label{fig:sub8}
    \end{subfigure}
    \begin{subfigure}[b]{0.3\textwidth}
        \centering
        \includegraphics[height=6cm]{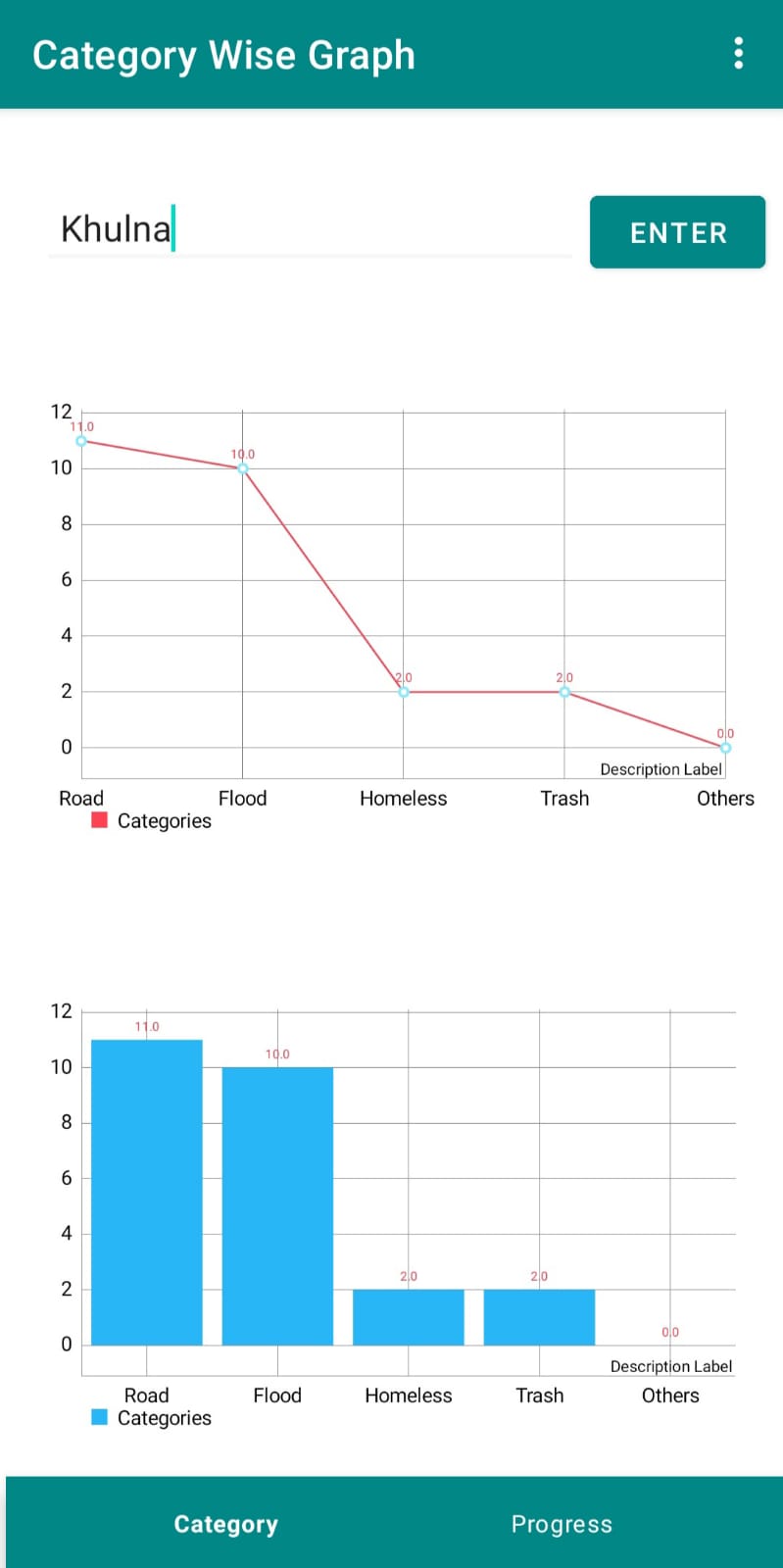}
        \caption{Statistics}
        \label{fig:sub9}
    \end{subfigure}
    
    \caption{Illustrative examples}
    \label{fig:overall}
\end{figure}

The system utilizes two separate Android applications to facilitate the reporting of municipal issues. The user version of the "CitySolution" application allows general citizens to register and submit photos of problems they are facing, along with any additional feedback. The location of the issue is automatically detected using Fused Location Services by Google. As depicted in Fig. \ref{fig:sub1}, a user submits a report through the application, capturing an image of a damaged road. Subsequently, the application automatically retrieves the location of the complaint, as illustrated in Fig. \ref{fig:sub2}. It is to be noted that, the country must be Bangladesh for submitting complaints. The image is then subjected to analysis by a deep learning model integrated within the user interface, resulting in the precise classification of the problem into one of the categories. Users can also see the history of their submitted complaints and also can see the details of the history illustrated in Fig. \ref{fig:sub3}. Users can view the problem status set by the city corporation or municipal authority. Initially, the status is set to "pending", but can be updated to "processing" and "solved" as the city corporation makes progress shown in Fig. \ref{fig:sub4}.

The second application is designed for the authorities and allows them to view and categorize complaints submitted by citizens. At registration, a QR code scanning is performed on this application to get the unique employee ID, name, and city. This feature ensures that only individuals with this QR code can register as employees. Since the employee ID must be unique for each registration, only one account can be opened with each QR code. Central admin can generate these QR codes through the app using the information of ID, first name, last name, and city. Authorities can use these QR codes to register. After registration and login, employees can see the precise location of the problem, update the problem status, and update the category (if the problem is misclassified), as well as manually identify any fake complaints shown in Fig. \ref{fig:sub5}. They can also see the precise location of the problem with the help of Google Maps which is illustrated in Fig. \ref{fig:sub6}. The authorities can also send feedback to the users, and the users will receive in-app notifications of any changes in the problem status shown in Fig. \ref{fig:sub7}. The higher authority has a separate panel with additional security measures during login, which allows them to monitor the overall system and verify registered city corporation members who use the app. They have the access rights to remove a registered employee. In this circumstance, the employee will get an email. The two applications also provide basic features such as user or employee profiles, language settings illustrated in Fig. \ref{fig:sub8} etc. 
Both users and higher authorities can view the number of solved, processing, and pending problems for any city, as well as the number of problems in different categories, in a graphical format provided in Fig. \ref{fig:sub9}. This ensures transparency at the municipal level.

\section{Impact}
In this system, the complaints of the users are classified into four categories- “Damaged Road”, ”Flood”, ”Trash” and “Homeless People”. The number of classes for the deep learning model can be increased in the future to encompass all possible problem aspects of a city. In the current system, the users must upload an image of the problem. A separate text-based segment can be added for written complaints and these texts may also be classified automatically. Furthermore, an automatic fake complaint detection mechanism can be incorporated.
This system propagates the development of a smart city by lessening the possibility of human error and solving the lack of automation, transparency, and central monitoring which existed in \cite{one} and \cite{two}.

Currently, there is a lack of nationwide applicability to online platforms for submitting citizen complaints, necessitating in-person complaint submission in certain districts. Furthermore, existing platforms typically require manual selection of complaint categories and locations. In contrast, these applications offer automated categorization, simplifying the reporting process for citizens who only need to capture and submit images of their complaints. Additionally, it establishes a communication link between city corporation officials and higher authority, promoting operational efficiency and transparency. Users also receive in-app notifications regarding the progress of their reported issues, fostering increased trust and dependability in city corporations. This, in turn, encourages city corporation employees to fulfill their duties. The platform also offers notable features such as automatic location detection, bilingual support, map, and email redirection to further enhance the system's functionality and user experience. The incorporation of various graphical representations within the application facilitates comprehensive tracking and evaluation of city corporation tasks by both city corporation employees and higher authority. The system sends notifications to registered users when they submit fake complaints, thereby safeguarding the integrity of the reporting system and preventing further instances of such behavior. Using these applications, both the users and the authorities can save time by making city corporations smarter by classifying the complaints.

The intended user groups for this system are the citizens and the authority of the city corporation. The municipal issues of each city are handled in such a way that the employees can only see their city, not the other cities. So, in this case, the intended users are the city corporation or municipal authorities. The central authority can monitor any city. In this case, the intended user is the central government. These applications only handle locations inside Bangladesh, and for this reason, the applications have been made available in both English and Bengali.

\section{Conclusions}
This paper aims to bridge the gap between the citizens and the authorities to ensure the development of a city. At its core lies the principle of 'Develop the City Together,' a call to action for increased citizen involvement, alongside a commitment to unwavering transparency and accountability from all parties involved. Using the power of deep learning, the methodology enables accurate problem classification and facilitates targeted solutions. Moreover, the user-friendly approach ensures accessibility for both the citizens and the city authorities, empowering collaboration towards a shared vision of progress and prosperity.

\section*{Acknowledgements}
\label{}
The authors would like to express their sincere gratitude to Khulna University of Engineering \& Technology for offering a suitable environment for developing these applications.




 \bibliographystyle{elsarticle-num} 
 \bibliography{references.bib}

\end{document}